\documentclass[%
 reprint,
 amsmath,amssymb,
 aps,
]{revtex4-2}

\usepackage[english]{babel} 
\usepackage{amssymb}
\usepackage{txfonts}
\usepackage{mathdots}
\usepackage{graphicx} 
 
\newcommand{\pd}[2]{\frac{\partial #1}{\partial #2}}

\newcommand{\ket}[1]{\left| #1 \right>} 
\newcommand{\braket}[2]{\left< #1 \vphantom{#2} \right|
 \left. #2 \vphantom{#1} \right>} 

\begin{document}
\title{ Quantum-Limited  Estimation of  Phase Gradient}

\author{Larson, W.}
\homepage{larsonwd@knights.ucf.edu}
\author{Saleh, B.E.A.}
\affiliation{CREOL, The College of Optics and Photonics, University of Central Florida, 4304 Scorpius St, Orlando FL, 32817}
\date{\today}
\begin{abstract}
We show that the quantum Cram\'er-Rao bound on the precision of measurements of the optical phase gradient, or the wavefront tilt, with a beam of finite width is consistent with the Heisenberg uncertainty principle for a single-photon state, and is a factor of 2 better for the two-photon state that is maximally entangled.  
This fundamental bound governs a trade-off between quantum sensitivity and spatial resolution.
Precision bounds based on a structured configuration using binary projective measurements implemented by an image-inversion interferometer, are higher, and the two-photon factor of 2 advantage is lost for large beam width or large phase gradient. In all cases, estimation of the phase gradient is compatible with estimation of the phase, allowing for optimal joint estimation of \emph{both} parameters simultaneously. 
\end{abstract}
\maketitle



\section{ Introduction}

The performance of classical optical metrology systems is limited by standard classical limits on the precision of measurement of optical phase and amplitude.  Quantum metrology is based on the use of optical probes in nonclassical states of light that enable precision superseding the classical limits; but new superior limits emerge as the ultimate quantum limits.  For example, the classical standard limit for estimation of the optical phase with a fixed mean number of photons $N$ is $1/{\sqrt{N}}$ \cite{Mandel1995OpticalOptics}, but the ultimate quantum limit for a fixed number of photons $N$ (Heisenberg limit) is $1/N$ \cite{Caves1981Quantum-mechanicalInterferometer,Braunstein1992,Holland1993,Simon2016,Giovannetti2004}.  

In this paper, we determine quantum limits on the precision of measurement of the optical \textit{phase gradient}, or the wavefront tilt, by use of an optical beam of finite width, assuming
single-photon and two-photon entangled quantum states.  The precision limits are the Cram$\acute{\mathrm{e}}$r-Rao bounds computed from the quantum Fisher information (QFI) \cite{Helstrom1976,Holevo1982,Braunstein1992,Braunstein1994StatisticalStates,Giovannetti2011,Giovannetti2006QuantumMetrology,Matsumoto2002,Demkowicz-Dobrzanski2012}.  We show that these limits are inversely proportional to the beam width, in accord with the Heisenberg uncertainty principle \cite{Sun2017UncertaintyEstimations} and its generalization to a spatially-coded two-photon state \cite{Braunstein1995,Shin2011QuantumMeasurements,Jin2017Quantum-enhancedCounting} that probes the phase slope in a manner similar to that of the states used in so-called "NooN-state" interferometery \cite{Dowling2008,Lee2004FromInterferometry,STREKALOV2002Two-photonImaging,Dorner2009}.  We also calculate limits on the precision of phase gradient measurements implemented by use of a structured configuration using binary projective field measurements in a scanning image-inversion interferometer \cite{larson2019,Tang2016,Cohen2014Super-resolvedMeasurement.}. 

Finally, we extend our analysis to the concurrent quantum estimation of both phase and phase gradient and establish "compatibility" between these parameters, i.e., that the optimal precision in the multiparameter measurement scheme equals that of the optimal precision for measurement of each parameter alone\cite{Ragy2016}.

\section{Quantum Fisher Information for single- and two-photon states}

 An optical beam probes a phase object that introduces a linearly varying phase $\varphi (x)=\varphi_0+\theta x$ in the plane orthogonal to the beam direction. The phase gradient $\theta$ is to be estimated by use of measurements on the transmitted beam. This model is applicable if the phase varies slowly within the area of the beam, in which case $\varphi_0 = \varphi(x=0)$ and $\theta =\partial \varphi /\partial x \,|_{x=0}$.   
 
 If the quantum state of the light transmitted through the phase object is $\ket{\psi} $, assumed to be a pure state, then the quantum Fisher information (QFI) is \cite{Demkowicz-Dobrzanski2015QuantumInterferometry,Fujiwara1995QuantumEstimation}
 \begin{equation}
F_Q(\theta)=4 \left( \braket{\psi'}{\psi'}-|\braket{\psi}{\psi'}|^2\right), 
\label{QFI_DefPure}
 \end{equation}
where $\psi'$ refers to the derivative of $\psi $ with respect to $\theta$. The quantum Cram\'er-Rao (QCR) bound on the variance of the estimate of $\theta$ is $\textrm{Var}(\theta) \geq {\sigma}^2_{\theta }=1/F_Q\left(\theta \right)$. In this section, we determine $F_Q(\theta)$ and the associated precision bound $\sigma_x$ for light in two cases: single-photon and two-photon pure quantum state.

\subsection{Single-photon state}
 If the single-photon state is a pure quantum state $\ket{\psi_0} =\int dx\ \psi_0(x)\ket{x}$, where $\psi_0(x)$ is an arbitrary wavefunction normalized such that $\int |\psi _0(x)|^2 \ dx=1$, then upon transmission through the phase object the state becomes 
 \begin{equation}
\ket{\psi} =\int dx \ e^{-i\theta x}{\psi }_0\left(x\right)\ket{x} .  
 \end{equation} 
 Based on Eq. \eqref{QFI_DefPure}, the QFI is
 \begin{equation}
F^{(1p)}_Q\left(\theta \right)=4\left\{\int dx{\ x}^2{\left|{\psi }_0\left(x\right)\right|}^2-{\left|\int dx\ x{\left|{\psi }_0\left(x\right)\right|}^2 \,\right|}^2\right\}.  \label{eq3}
 \end{equation}
 If $\psi _0\left(x\right)$ is an even function, then the second term in \eqref{eq3} vanishes, and 
\begin{equation}
F^{\left(1p\right)}_Q\left(\theta \right)=4{\sigma }^2_x,
\label{QFI_OneP}
\end{equation}
where $\sigma^2_x=$  $\int dx{\ x}^2\left|\psi_0\left(x\right)\right|^2$ is the second moment of the probability density function $\left|\psi_0\left(x\right)\right|^2$ and $\sigma_x$ is a measure of its width.  The QCR bound on the variance of the estimate of $\theta$ is ${\sigma}^2_{\theta }=1/F_Q\left(\theta \right)$, so that \begin{equation}
{\sigma }_x{\sigma }_{\theta }=\frac{1}{2}.
\end{equation}
Because the phase gradient $\theta $ equals the transverse component $q$ of the wavevector, this is simply an expression of the bound dictated by the Heisenberg uncertainty principle $\sigma_x\sigma_q=\frac{1}{2}$. 

\subsection{Two-photon state}
A two-photon pure quantum state is described by the integral 
$\ket{\psi_0}=\iint dx_1 \ dx_2 \ \psi_0(x_1,x_2)\ket{x_1,x_2}, $
where $\psi_0(x_1,x_2)$ is an arbitrary two-photon wavefunction normalized such that
$\iint{{dx_1 \ dx_2\ \left|\psi_0(x_1,x_2)\right|}^2=1}.$ 
Upon transmission through the phase object, the state becomes 
\begin{equation}
\ket{\psi}=\iint   dx_1 \, dx_2 \ \psi_0\left(x_1,x_2\right)e^{-i\theta \left(x_1+x_2\right)}\ket{x_1,x_2}.
\label{eq:two-photon-state}
\end{equation}
Using Eq. \eqref{QFI_DefPure}, the QFI is  
\begin{multline}
F^{\left(2p\right)}_Q\left(\theta \right)=
4 \iint dx_1dx_2 \left(x_1+x_2\right)^2{|{\psi }_0(x_1,x_2)|^2}
\\
-4{\left|\iint{dx_1dx_2}\left(x_1+x_2\right){{|\psi }_0\left(x_1,x_2\right)|^2}\, \right|}^2.  \label{eq-FQ2p}
\end{multline}
Assuming a maximally entangled state $\psi_0x_1,x_2)=f_0(x_1)\delta(x_1-x_2)$ \cite{Schneeloch2016IntroductionZone} and that $f_0(x)$ is an even function, the second term of Eq.(\ref{eq-FQ2p}) vanishes and the QFI is
\begin{equation}
F^{\left(2p\right)}_Q\left(\theta \right)=16\,{\sigma }^2_x,
\label{QFI_TwoP}
\end{equation}
where $\sigma^2_x=\int dx\ x^2{\left|f_0\left(x\right)\right|}^2$ is a measure of the width of ${\left|f_0\left(x\right)\right|}^2.$  Therefore, the minimum uncertainty ${\sigma }_{\theta }$ of estimates of the phase-gradient satisfies the relation
 \begin{equation}
{{\sigma }_x\sigma }_{\theta }=\frac{1}{4}.  \label{eq:product}
\end{equation}

The bound for the entangled two-photon Heisenberg uncertainty product is therefore smaller by a factor of 2 than that of the single-photon case, assuming equal widths of the functions ${\left|{\psi }_0\left(x\right)\right|}^2$ in the single-photon case and ${\left|f_0\left(x\right)\right|}^2$ in the two-photon case. 

Note that if the two-photon state $\psi(x_1,x_2)$ is separable, then \eqref{eq-FQ2p} yields $F^{\left(2p\right)}_Q\left(\theta \right)=8\,{\sigma }^2_x$, so that ${{\sigma }_x\sigma }_{\theta}=\tfrac{1}{2\sqrt{2}}$, which is what would be obtained for estimates based on two independent single-photon realizations.  

To show that the maximally entangled state yields the highest QFI, we consider the wave function $\psi(x_1,x_2)$ in the coordinates $x_-= x_1-x_2$ and   $x_+= (x_1+x_2)$ and assume even symmetry in these coordinates. Equation (\ref{eq-FQ2p}) then yields Eq. (\ref{QFI_TwoP}), where $\sigma_x$ is replaced by the width $\sigma_{x+}$ of $\psi(x_1,x_2)$ along the $x_+$ coordinate. Since $\psi(x_1,x_2)$ has unit norm, the product of its widths $\sigma_{x+}$ and $\sigma_{x-}$ along $x_+$ and $x_-$ is fixed, so that maximum  QFI is obtained with the largest $\sigma_{x+}$ and the smallest $\sigma_{x-}$, which leads in the limit to the maximally entangled state.

\section{ Phase Gradient Estimation With Image-Inversion Interferometers}

We now consider specific configurations for measuring the phase gradient and assess their expected precision in comparison with the ultimate quantum bounds described by \eqref{QFI_OneP} and \eqref{QFI_TwoP}. We will also find bounds on the precision associated with single-photon and two-photon quantum states.

\begin{figure}
    \centering
    \includegraphics[scale=0.85]{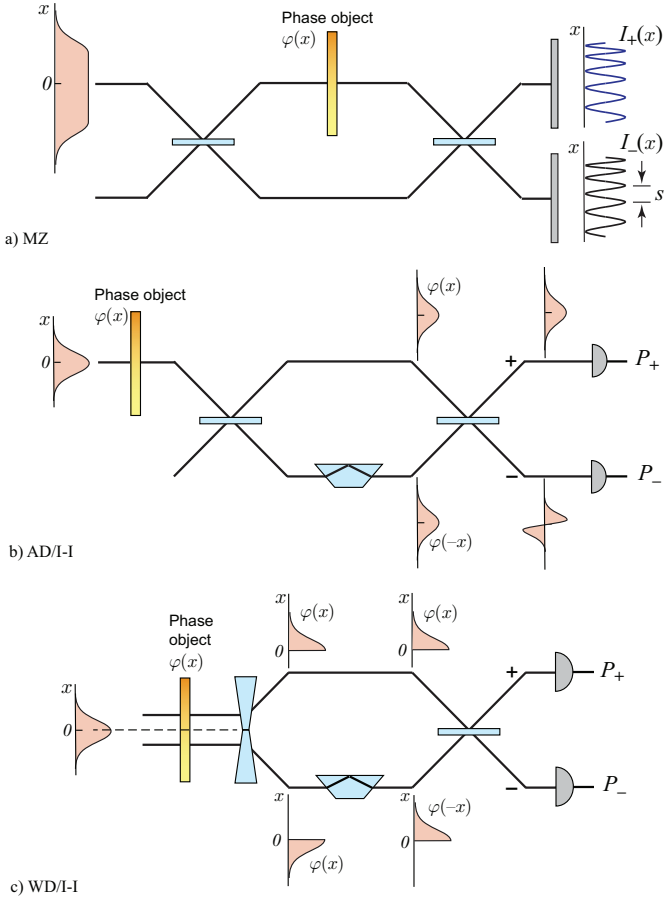}
    \caption{Configurations for measurement of the phase gradient. (a) Measurement of fringe spacing $s$ in a conventional MZ interferometer. (b) Measurement of the optical powers $P_+$ and $P_-$ at the output ports of a scanning image-inversion interferometer, which separates the even and odd components of an even illumination modulated by the phase object. (c) Wavefront-division image-inversion interferometer. }
    \label{fig:Diagram Setup}
\end{figure}

 \textbf{i) Mach Zehnder Interferometer}. 
The first configuration is the Mach Zehnder (MZ) interferometer, which is commonly used to measure an optical phase. The phase object is placed in one arm, with the other arm empty, and a planar wave or expanded beam is used to generate an interference pattern, as illustrated  in Fig.\ref{fig:Diagram Setup}(a). The phase object modulates the planar wave by a factor $e^{i\theta x}$, which bends it by an angle 
$\alpha ={{\mathrm{sin}}^{\mathrm{-}\mathrm{1}} \left(\lambda /2\pi \theta \right)}$,  where $\lambda$ is the wavelength.  Upon interference with the unmodulated planar wave, the outcome is an interference pattern with fringe spacing  $s=\lambda /{\sin{\alpha}}=2\pi /\theta $.  The phase gradient $\theta $ at $x$ may be estimated by measuring the fringe spacing in the neighborhood of $x$.    

 \textbf{ii) Image-Inversion Interferometer}. In the second configuration, the beam modulated by the phase object is interrogated by an MZ interferometer with an extra mirror in one arm, as illustrated in Fig.1(b). The system acts as an image-inversion (I-I) interferometer \cite{Weigel2015a,Weigel2015AberrationInterferometer}. 
For an optical beam of amplitude  ${\psi }_o\left(x\right)$ and width ${\sigma }_x$, the beam transmitted through (or reflected from) the phase object has an amplitude $\psi \left(x\right)={\psi }_o\left(x\right)e^{i\varphi \left(x\right)}$ which is mixed with an inverted copy of itself $\psi \left(-x\right)$ to generate amplitudes ${{\frac{1}{2}}}\left[\psi \left(x\right)\pm \psi \left(-x\right)\right]\ $at the output ports of the interferometer. The interferometer can be made using spatially-separated paths, as we depict in Fig.\ref{fig:Diagram Setup}(b) or another ancillary degree of freedom such as polarization \cite{Aiello2015a,Nair2016a,Weigel2015AberrationInterferometer,larson2019}.  The corresponding intensities $I_{\pm }\left(x\right)=\frac{1}{2}\left|\psi \left(x\right)\pm \psi \left(-x\right)\right|^2\ $are measured with two detectors of areas greater than the beam cross-section ${\sigma }_x$.  The result is the two signals   $p_{\pm }=\frac{1}{2}\pm \frac{1}{2}{\textrm{ Re}}\int
{dx{\ \psi }^*_0\left(x\right){\psi }_0\left(-x\right)e^{i2\theta x},}$ where we have assumed that $\int
{{\left|{\psi }_0\left(x\right)\right|}^2dx=1.}$ In essence, this binary measurement represents a projection of the spatial distribution of the photon onto its even ($+$) and odd ($-$) components. 

If ${\psi }_0\left(x\right)$ is an even function, then 
\begin{equation}
\begin{aligned}
P_+ =\int
\, dx \ |\psi_0\left(x\right)|^2\cos^2 \left(\theta x\right),\\ 
P_- =\int
\, dx \  |\psi_0\left(x\right)|^2\sin^2 \left(\theta x\right).
\label{SinglePProbIntegrals}
\end{aligned}
\end{equation}

\noindent For example, for a Gaussian function $\left|\psi_0 \left(x\right)\right|^2= (1/\sqrt{2\pi}\sigma_x)\exp(-x^2/2\sigma_x^2 )$,
\begin{equation}
P_{\pm }=\frac{1}{2}\left(1\pm e^{-2{\theta }^2{\sigma}^2_x}\right).
\label{SinglePhotonProbabilities}
\end{equation}

 The difference between these signals, $P_+-P_-=\exp \left({-2{\theta}^2{\sigma }^2_x}\right)$, is a decreasing function of the beam width $\sigma_x$ extending over a range of $\sigma_x \theta$ from 0 to 2, so that the phase gradient $\theta$ can be readily discerned in this range.   In contrast, in the MZ interferometer the beam width must be larger than the fringe spacing, i.e.,  $\sigma_x > s=2\pi /\theta $, or $\sigma_x \theta > 2\pi$. Therefore, the beam width in the scanning I-I interferometer can be much smaller than that used in the MZ interferometer, offering greater resolution when imaging the spatial variation of $\theta$.   
 
 \textbf{iii) Wavefront-Division Image-Inversion Interferometer}.
 Another rendition of the I-I interferometer uses wavefront division (WD), as depicted schematically in Fig.\ref{fig:Diagram Setup}(c). The beam field $\psi_0(x)$ is split into two spatial modes: the positive spatial mode $\psi_+\left(x\right)=\psi_0\left(x\right),\ x>0$, and zero otherwise; and the negative spatial mode ${\psi }_-\left(x\right)=\psi_0\left(x\right),\ x<0$, and zero otherwise. Upon transmission through the phase object, both modes are modulated by the object phase $\varphi(x)$. After image inversion, the negative mode is converted into a positive mode, but with phase $\varphi(-x)$, so that the phase difference between the modes becomes $\varphi(x)-\varphi(-x)=\theta x$. When the modes are combined by a beam splitter and detected, the measured powers are given by the same equation (\ref{SinglePProbIntegrals}) as in the conventional I-I interferometer, which uses amplitude division (AD) interferometry.   Wavefront division may be implemented by use of prisms or spatial light modulators (SLM) that separate the spatial modes into different optical paths. Alternatively, a polarization-sensitive SLM may be used to convert the spatial modes into polarization channels \cite{larson2019}.  

\subsection{Single-photon state}
 \noindent \textbf{i) Image-Inversion Interferometer}. 
 We first consider the I-I interferometer shown in Fig.\ref{fig:Diagram Setup}(b) or (c). If the probe wave is in a single-photon state, then the above classical analysis is applicable with the signals $P_+\ $and $P_-$ interpreted as the probabilities of the photon being detected in the $+$ and $-$ output ports, respectively. The Fisher information associated with such measurement is 
 \begin{equation}
F^{\left(1p\right)}\left(\theta \right)=\frac{1}{P_+}{\left(\frac{dP_+}{d\theta }\right)}^2+\frac{1}{P_-}{\left(\frac{dP_-}{d\theta }\right)}^2.  \label{eq:I-IF1p}
\end{equation}
Using the expressions in \eqref{SinglePhotonProbabilities}, it follows that 
 \begin{equation}
F^{\left(1p\right)}\left(\theta \right)=4{\sigma}^2_x \,/\, \zeta^2  \left({\sigma }_x\theta \right),\end{equation}
where 
$\zeta^2 \left(y\right)= \sinh{(2y^2)}/2y^2$
is a monotonically increasing function of $y$ with value equal to 1 for $y\to 0.$   Therefore, in the limit ${\sigma }_x\theta \ll 1,$ i.e., when the phase varies slowly within the beam width,  the factor $\zeta \left({\sigma }_x\theta \right)=1,$ so that $F^{\left(1p\right)}\left(\theta \right)=F^{\left(1p\right)}_Q\left(\theta \right)$, i.e., the image-inversion interferometer provides the best possible precision for estimating $\theta $.  Fig. \ref{fig:FisherInformationComparisons} illustrates the dependence of the Fisher information $F^{\left(1p\right)}\left(\theta \right)$ on the beam width ${\sigma }_x$.  The Cram\'er-Rao estimation error ${\sigma }_{\theta }$ corresponding to $F^{\left(1p\right)}\left(\theta \right)$ satisfies the relation 
\begin{equation}
{\sigma }_x{\sigma }_{\theta }=\frac{1}{2}\ \zeta \left({\sigma }_x\theta \right).
\end{equation}
As illustrated in Fig.\ref{fig:UncertaintyProductComparisons} the uncertainty product equals $\frac{1}{2}$ for small ${\sigma }_x\theta $ and increases monotonically with increase of ${\sigma }_x\theta $. 

\vspace{5pt}

 \textbf{ii) Mach-Zehnder Interferometer}.  With a single-photon source at the input of the conventional MZ interferometer of Fig.\ref{fig:Diagram Setup}(a), the probability distributions of the photon being detected at position $x$ in the upper and the lower arms are
\begin{equation}
\begin{aligned}
    p_+(x)= |\psi_0(x)|^2\cos^2 (\theta x/2), \\
    p_-(x)= |\psi_0(x)|^2\sin^2 (\theta x/2).
\end{aligned}
\end{equation}
The derivatives of these probabilities with respect to $\theta$ are 
\begin{equation}
    p'_+(x)=-\cos (\theta x) \sin(\theta x)\, x\, |\psi(x)|^2=-p'_-(x).
\end{equation}
Substituting in Eq. (\ref{eq:I-IF1p}), the total Fisher information is 
\begin{equation}
    F^{(1p)}(\theta)=
    \sigma_x^2,
\end{equation}
where we have assumed a perfect detector.  
 Thus the Fisher information for the MZ interferometer is therefore a factor of $4$ smaller than the QFI $F_Q^{(1p)}(\theta)$ (see Eq. (\ref{QFI_OneP})). This factor originates from the fact that only half of the optical power, on average, probes the phase object when compared to the I-I interferometer, wherein the beam paths are split after transmission through the phase object. 
 
\subsection{Two-photon state}
 For light in the two-photon phase-modulated quantum state given by Eq. (\ref{eq:two-photon-state}), and assuming maximal entanglement, 
 \begin{equation}
 \ket{\psi} =\int dx \ f_0\left(x\right)e^{-i2\theta x}\ket{x,x}. \label{eq-Initial-State}
 \end{equation}
To take advantage of entanglement we use the WD/I-I interferometer shown in Fig.\ref{fig:Diagram Setup}(c), which has a spatially selective element that splits the phase-modulated incoming two-photon wavefront into two spatial modes: $\ket{+}$ and $\ket{-}$.   
 The result is a superposition state    
 \begin{equation*}
 \int dx \ e^{-i2\theta x} \big[ \psi_+\left(x\right) \ket{+,+} 
 + \psi _-\left(x\right)\ket{--} \big]\, \ket{x,x}.    
 \end{equation*}
 With one of the two channels, say $\ket{-} $, introducing image inversion, the state becomes 
 \begin{equation*}
 \int dx \,
 \big[ e^{-i2\theta x} \ket{+,+}  +  \ e^{+i2\theta x} \ket{-,-} \big]\\
\psi_+\left(x\right) \ket{x,x}.  
 \end{equation*} 
 The two channels are then combined with a regular beam splitter to produce the state
\begin{equation*}
      \int  dx \, \big[  \cos(2\theta x) \ket{c} +i\sin(2\theta x) \ket{a} \big] \psi_+(x)\ket{x,x},
\end{equation*}
where $\ket{c}=\frac{1}{\sqrt{2}}\left(\ket{+,+}+\ket{-,-}\right)$ and $\ket{a}=\frac{1}{\sqrt{2}}\left(\ket{+,-} +\ket{-,+}\right)$ 
 are correlated and anticorrelated states, respectively. It follows that the probability $p_a=p(1,1)$ of measuring one photon in each channel (anti-correlated outcome) and the probability $p_c=p(2,0)+p(0,2)$ of measuring the two photons together in either channel (correlated outcome) are: 
\begin{equation}
\begin{aligned}
p_{c}=&\int dx \, |\psi_0(x)|^2\cos^2 \left(2\theta x\right),\\
p_{a}=&\int dx\, |\psi_0(x)|^2\sin^2 \left(2\theta x\right).
\end{aligned}
\end{equation}

Since these expressions are identical to those in Eq. (\ref{SinglePProbIntegrals})  for the single-photon case, except for $\theta$ being replaced by $2\theta $, it follows that if $|f_0\left(x\right)|^2$ is a Gaussian function 
$(1/\sqrt{2\pi}\sigma_x)\exp(-x^2/2\sigma_x^2),$
the Fisher information is 
\begin{equation}
F^{\left(2p\right)}\left(\theta \right)=16{\sigma }^2_x \, / \, \zeta^2 \left({4\sigma }_x\theta \right)=4F^{\left(1p\right)}\left(2\theta \right).
\end{equation}

\begin{figure}
    \centering
    \includegraphics[scale=0.7]{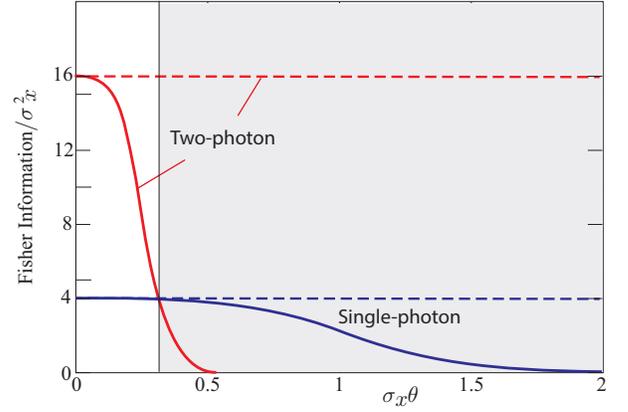}
    \caption{Fisher information for estimation of the phase gradient $\theta$ using an optical beam of width $\sigma_{x}$ in a single-photon state (blue) and a two-photon state (red). Dashed lines indicate the ultimate bound  dictated by the quantum Fisher information. Solid lines indicate Fisher information for a binary measurement configuration using an image-inversion interferometer.  For $ \sigma_x\theta \geq 0.32$ (shaded area), the two-photon state loses its advantage over the single-photon state.}
    \label{fig:FisherInformationComparisons}
\end{figure}

\begin{figure}
    \centering
    \includegraphics[scale=0.7]{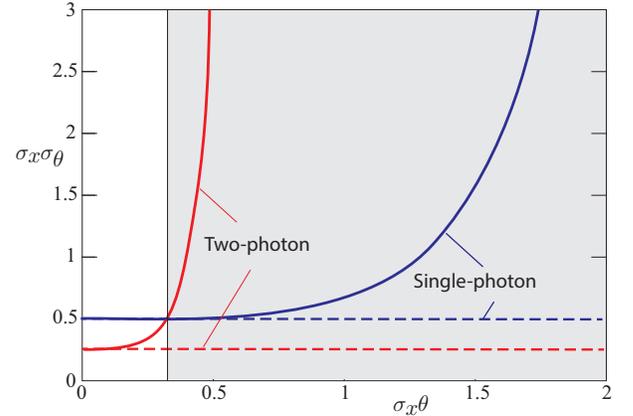}
    \caption{Minimum uncertainty product $\sigma_x\sigma_\theta$ versus $\sigma_x\theta$ for the single-photon state (blue) and the two-photon state (red). Dashed lines indicate the ultimate bound dictated by the quantum Fisher information. Solid lines are based on Fisher information for a binary measurement configuration using an image-inversion interferometer.  For $ \sigma_x \theta\geq 0.32$ (shaded area), the two-photon state is no longer providing lower uncertainty product. }
    \label{fig:UncertaintyProductComparisons}
\end{figure}

As illustrated in Fig.\ref{fig:FisherInformationComparisons}, for small ${\sigma }_x\theta$, the maximum attainable Fisher information is a factor of four greater than that in the single photon case, but it decreases at a rate four times greater than the single photon case, as ${\sigma }_x\theta $ increases.  The uncertainty product in the two-photon case is 
\begin{equation}
{{\sigma }_x\sigma }_{\theta }=\frac{1}{4} \, \zeta \left(4{\sigma }_x\theta \right). 
\end{equation}
A plot of the uncertainty product for the single- and two-photon cases is shown in Fig.\ref{fig:UncertaintyProductComparisons}. For small ${\sigma }_x\theta ,$ the two-photon uncertainty product is a factor of 2 smaller, but for ${\sigma }_x\theta >0.3199,$ it is greater.    The two graphs intersect at ${\sigma }_x\theta =0.3199$

\section{Concurrent estimation of phase and phase gradient}
In this section, we determine bounds on the precision of concurrent estimates of both the phase $\varphi_0$ and the phase gradient $\theta$ of a phase object $\varphi(x)=\varphi_0+\theta x$ at $x=0$, treated as a multiparameter estimation problem \cite{Ragy2016,Roccia2018MultiparameterVisibility,Pezze2017OptimalPhases,Szczykulska2016}. Evidently, estimation of $\varphi_0$ requires the use of a reference, i,e, embedding the object in an interferometer, such as the MZ interferometer in Fig.\ref{fig:Diagram Setup}(a). 
In this case, the QFI may be based on the optical fields in both arms, either before or after the second beam splitter. For simplicity, but without loss of generality, we calculate the QFI matrix for the optical field after the second beam splitter. 
This allows us to calculate the multiparameter QCR bounds, which will set the best sensitivity with which $\varphi_0$ and $\theta$ may be estimated. 

We also design a structured configuration for measurement of $\varphi_0$ and $\theta$ that uses I-I interferometers in the output ports of the MZ interferometer, and show that in the limit of small $\sigma_x \theta$, the structured configuration provides measurement sensitivity equal to the QCR bound.

\subsection{Single-Photon State}
\textbf{Quantum Fisher information.}
The state of single-photon light at the output ports of the MZ interferometer in Fig.\ref{fig:Diagram Setup}(a) is
\begin{multline}
\ket{\psi} =\int dx \, \big[\cos(\tfrac{1}{2}(\varphi_0+x\theta))\ket{T}\\
+i\sin(\tfrac{1}{2}(\varphi_0+x\theta))\ket{B}\big]\,
\psi_0\left(x\right)\ket{x}, 
\label{MultiParamState}
 \end{multline}
 where $\ket{T}$ and $\ket{B}$ represent the upper and lower output ports of the MZ interferometer. The derivative of this state with respect to $\varphi_0$ and $\theta$ are  
 \begin{multline}
\ket{\psi_{\varphi_0}}= \tfrac{1}{2}\int dx \, 
\big[ -\sin(\tfrac{1}{2}(\varphi_0+x\theta))\ket{T}\\
+i\cos(\tfrac{1}{2}(\varphi_0+x\theta))\ket{B}\big]\, 
\psi_0\left(x\right)\ket{x},   \label{eq-psi-phi}
 \end{multline} 
 and  
  \begin{multline}
\ket{\psi_{\theta}}= \tfrac{1}{2}\int dx \, 
\big[ -\sin(\tfrac{1}{2}(\varphi_0+x\theta))\ket{T}\\
+i\cos(\tfrac{1}{2}(\varphi_0+x\theta))\ket{B} \big]\,
x\, \psi_0\left(x\right)\ket{x},  \label{eq-psi-theta}
 \end{multline} 
respectively.
The multiparameter QCR bounds are 
\begin{equation}
\sigma^2_{\varphi_0}=\left[ \textbf{F}_Q\left(\theta,\varphi_0 \right) \right]^{-1}_{\varphi_0\varphi_0}, \quad
\sigma^2_{\theta}=\left[ \textbf{F}_Q\left(\theta,\varphi_0 \right) \right]^{-1}_{\theta\theta},
\end{equation}
where the elements of the QFI matrix  $\textbf{F}_Q$ are  

 \begin{equation}
 \left[\textbf{F}_Q(\theta,\varphi_0)\right]_{i,j}
 =4\textrm{Re}\left\lbrace
 \braket{\psi_i}{\psi_j}- 
 \braket{\psi}{\psi_i} \braket{\psi}{\psi_j}  \right\rbrace, 
\label{QFIM_Def}
 \end{equation}
 and $\ket{\psi_j}$ is the partial derivative of $\psi$ with respect to the parameter $j$. 
 
Based on Eqs. (\ref{eq-psi-phi}) and (\ref{eq-psi-theta}), the QFI matrix has diagonal elements 
$\braket{\psi_{\varphi_0}}{\psi_{\varphi_0}}=1/4$ and
$\braket{\psi_{\theta}}{\psi_{\theta}}=\sigma_x^2/4$,
while the off-diagonal elements $\braket{\psi_{\theta}}{\psi_{\varphi_0}}=0$. i.e., the two derivatives of the state are orthogonal.
Since the QFI matrix is diagonal, there exists some measurement for which both $\theta$ and $\varphi_0$ can be concurrently estimated with QFI 
\begin{equation}
    \left[ \textbf{F}_Q\right]_{\varphi_0 \varphi_0}=1, \hspace{10pt} \left[ \textbf{F}_Q\right]_{\theta \theta}=4\sigma_x^2.
\end{equation}

Hence, the  QCR bounds are $\sigma_{\varphi_0}=1$ and $\sigma_\theta = 1/\sigma_x$. This establishes compatibility between the two estimated parameters \cite{Ragy2016}, i.e., that it is possible to estimate both $\theta$ and $\varphi_0$ at the QCR bound without the indeterminacy of one parameter affecting the other. 

\textbf{Image-Inversion Interferometer.}
It has been shown before both in theory and in experiment that photon-counting measurements at the output ports of the MZ interferometer are sufficient to achieve the QCR bounds for measurements of $\varphi_0$ \cite{Caves1981Quantum-mechanicalInterferometer,Ben-Aryeh2012PhaseInterferometer,Okamoto2012,Okamoto2008}. 
We describe below an interferometric configuration for concurrently measuring both $\varphi_0$ and $\theta$  with  precision equal to the precision attained by separate measurements.  

\begin{figure}[b]
    \centering
    \includegraphics[scale=0.72]{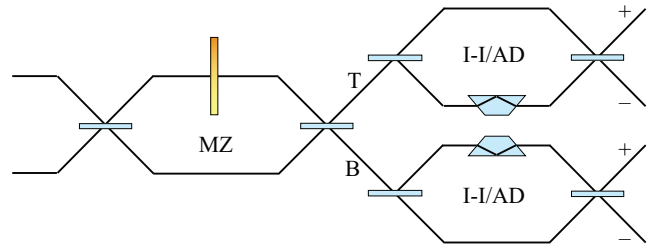}
    \caption{Configuration for concurrent estimation of the phase and the phase gradient using cascades of a Mach-Zehnder (MZ) interferometer and image-inversion (I-I) interferometers.}
    \label{fig:fig4}
\end{figure}

The configuration we consider, depicted in Fig.\ref{fig:fig4}, adds an \mbox{I-I} interferometer to each  of the output ports of an MZ interferometer. The composite system has four possible outcomes: the combinations of top and bottom MZ interferometer ports ($T$ or $B$) and the plus and minus ($+$ or $-$) of the I-I interferometer output modes.  

We treat the problem using the same single-photon model used in the calculation of the QFI matrix, assuming that a single photon enters one of the input ports of the MZ interferometer, and that we measure its arrival at any of the four output ports of the I-I interferometers. From Eq. (\ref{MultiParamState}), the associated probabilities are:
\begin{equation}
\begin{aligned}
P_{T+}&=\int dx \ |\psi_0(x)|^2\cos^2 (\varphi_0/2) \cos^2(x\theta /2) \ket{x},\\     
P_{B+}&=\int dx \ |\psi_0(x)|^2\sin^2 (\varphi_0/2) \cos^2(x\theta /2)\ket{x},\\
P_{T-}&=\int dx \ |\psi_0(x)|^2\cos^2 (\varphi_0/2) \sin^2(x\theta /2) \ket{x},\\    
P_{B-}&=\int dx \ |\psi_0(x)|^2\sin^2 (\varphi_0/2) \sin^2(x\theta /2)\ket{x}.
\end{aligned}
\end{equation}

The elements of the multiparameter Fisher infromation matrix $\textbf{F}$, in contrast to the QFI matrix  $\textbf{F}_Q$, are  

\begin{equation}
    \left[\textbf{F}(\theta,\varphi_0) \right]_{ij}=\sum_k \  \pd{}{i}\ln(P_k)\pd{}{j}\ln(P_k),
\end{equation}
where $k=T_+, T_-, B_+, B_-$ and $i,j$ refer to $\theta$ and $\varphi_0$. 
If the estimation is instead based on the probability sums
$P_T=P_{T+}+P_{T-}$ and $P_B=P_{B+}+P_{B-}$, which are independent of $\theta$, and the probability sums
\begin{equation}
\begin{aligned}
    P_+=P_{T+}+P_{B+}=\int^{\infty }_{-\infty} \ dx \ |\psi_0\left(x\right)|^2\cos^2 \left(x\theta /2\right),\\ 
    P_-=P_{T-}+P_{B-}=\int^{\infty }_{-\infty }\ dx \  |\psi_0\left(x\right)|^2\sin^2 \left(x\theta /2\right), \label{eqSum}
\end{aligned}
\end{equation}
which are independent of $\varphi_0$, it becomes evident that the Fisher matrix is diagonal, and because of the similarity of Eq. (\ref{eqSum}) to 
Eq. (\ref{SinglePProbIntegrals}), the diagonal elements are
 \begin{equation}
F^{\left(1p\right)}\left(\varphi_0 \right)= 1, \quad F^{\left(1p\right)}\left(\theta \right)=4{\sigma}^2_x \,/\, \zeta^2  \left({\sigma }_x\theta /2 \right).
\end{equation}
Therefore, the structured configuration of Fig.4 attains the QCR bounds for separate measurements of $\theta$ and $\varphi_0$. 

\subsection{Two-photon State}

\textbf{Quantum Fisher Information.}
Using an MZ interferometer with a single photon at each of its input ports, the output state is
\begin{equation}
 \ket{\psi^{(2p)}}=\ket{\psi_C} + i\ket{\psi_A}  \label{2pmultipstate0}
\end{equation}
where 
\begin{align}
        \ket{\psi_C} &=\int dx \ f_0(x) \cos(\varphi_0+\theta x) \ket{C} \ket{x,x}\\
        \ket{\psi_A} &=\int dx \ f_0(x) \sin(\varphi_0+\theta x) \ket{A} \ket{x,x},
    \label{2pmultipstate}
\end{align}
   $\ket{C}=\tfrac{1}{\sqrt{2}}(\ket{T,T}+\ket{B,B})$ is the correlated state, and $\ket{A}=\tfrac{1}{\sqrt{2}}(\ket{T,B}+\ket{B,T})$ is the anti-correlated state.
The derivatives of this state with respect to $\theta$ and $\varphi_0$ are  
\begin{multline}
    \ket{\psi^{(2p)}_{\theta}}=\int dx \ x \ f_0(x) \ket{x,x} \\
    \big{[} i\cos(\varphi_0+\theta x)\ket{A} -\sin(\varphi_0+\theta x) \ket{C}\big{]} 
    \label{thetaprime_2pmultipstate}
\end{multline}
and
\begin{multline}
    \ket{\psi^{(2p)}_{\varphi_0}}=\int dx \ f_0(x) \ket{x,x} \\
    \big{[} i\cos(\varphi_0+\theta x)\ket{A}  -\sin(\varphi_0+\theta x) \ket{C} \big{]}.
    \label{phiprime_2pmultipstate}
\end{multline}
Since the off-diagonal elements $\braket{\psi_{\theta}}{\psi_{\varphi_0}}=0$, the QFI matrix is diagonal, and hence there exists some measurement for which  $\varphi_0$ and $\theta$  can be independently estimated with the optimal precision for their joint estimation.  The diagonal elements are
$\braket{\psi_{\varphi_0}}{\psi_{\varphi_0}}=1$ and
$\braket{\psi_{\theta}}{\psi_{\theta}}=\sigma_x^2$,
so that $\left[ \textbf{F}_Q\right]_{\varphi_0 \varphi_0}=4,$ and $\left[ \textbf{F}_Q\right]_{\theta \theta}=16 \,\sigma_x^2$, and hence  
\begin{equation}
    F^{\left(1p\right)}_Q\left(\varphi_0 \right)=4, \hspace{10pt} F^{\left(1p\right)}_Q\left(\theta \right)=16\,{\sigma }^2_x.  \label{eq-QFI-1p}
\end{equation}
Again, the MZ interferometer offers the factor of 2 advantage associated with the two-photon state \cite{Hofmann2009AllInterferometry,Seshadreesan2013PhaseDetection,C.K.Hong1987}. 

\begin{figure}[b]
    \centering
    \includegraphics[scale=0.7]{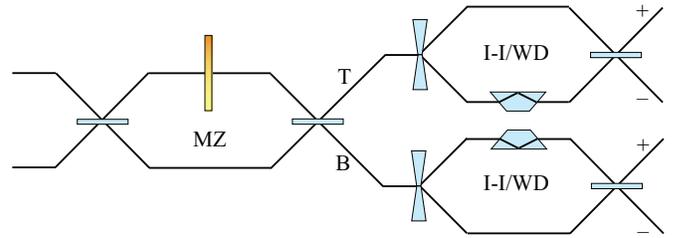}
    \caption{Configuration for concurrent estimation of the phase and the phase gradient using cascades of a Mach-Zehnder (MZ) interferometer and two wavefront-division image-inversion (WD/I-I) interferometers.}
    \label{fig:fig5}
\end{figure}

\textbf{Image-Inversion Interferometer.}
The configuration we use for concurrent estimation of phase and phase gradient with entangled two-photon light is a combination of an MZ interferometer with an WD/I-I interferometer attached to each of its output ports, as shown schematically in Fig.\ref{fig:fig5}.  This configuration is designed to enable separate measurements producing outcomes that are sensitive to only the phase and others that are sensitive to only the phase gradient.  With the two-photon state in Eq. (\ref{eq-Initial-State}) at the input of the MZ interferometer, and accounting for the 16 possibilities for the two photons arriving at the four output ports, we will now show that the Fisher information matrix is again diagonal, and the diagonal elements yield the Fisher information
\begin{equation} 
F^{\left(2p\right)}\left(\varphi_0 \right)=4, \hspace{10pt} 
F^{\left(2p\right)}\left(\theta \right)=16\,{\sigma}^2_x \,/\, \zeta^2 \left({\sigma }_x\theta \right),
\label{eq-FI-2p}
\end{equation}
which are identical to the QFI for measurements of $\varphi_0$ and $\theta$ independently. It is thus evident that this measurement scheme provides the full advantage of the two-photon state for the phase gradient $\theta$, even while estimating the phase $\varphi_0$.
 
The proof begins with the state at the output of the MZ interferometer in Eqs.(\ref{2pmultipstate0}) and (\ref{2pmultipstate}) and follows the correlated state $\ket{\psi_C}$ and the anticorrelated state $\ket{\psi_A}$ through the system, one at a time.

The anticorrelated state
\begin{equation}
    \ket{\phi_{A}}=\int dx f_0(x)
    \sin(\varphi_0+\theta x)\ket{+,+}\ket{A}\ket{x,x}.
\end{equation}
is split by the wavefront division element into the $\ket{+}$ and $\ket{-}$ modes of the WD/I-I interferometers, becoming the state
\begin{equation}
\int dx \, \big{[} f_{+}(x)  \ket{+,+} + f_{-}(x) \ket{-,-}\big{]}\sin(\varphi_0+\theta x)\ket{A}\ket{x,x}, 
\end{equation}
and when combined by the final beam splitters it becomes
\begin{multline}
    \ket{\psi_{A,\mathrm{out}}}= \int dx \ket{A}f_{+}(x)\ket{x,x} \\  
\big{[} \sin \varphi_0\cos(\theta x) \ket{c} -\cos\varphi_0 \sin(\theta x)\ket{a} \big{]}, 
\end{multline}
where $\ket{c}=\tfrac{1}{\sqrt{2}}(\ket{+,+}+\ket{-,-})$ and
$\ket{a}=\tfrac{1}{\sqrt{2}}(\ket{+,-}+\ket{-,+})$.
Similarly, the output state generated by the correlated component $\ket{\psi_A}$ is  
 \begin{multline}
    \ket{\psi_{C,\mathrm{out}}}= \int dx   \ket{C} f_{+}(x)\ket{x,x} \\
\big{[} \cos \varphi_0 \cos(\theta x) \ket{c} + \sin \varphi_0 \sin(\theta x)\ket{a} \big{]}.
\end{multline}
From the total state $\ket{\psi^{(2p)}_{\mathrm{out}}}=\ket{\psi_{C,\mathrm{out}}}+\ket{\psi_{A,\mathrm{out}}}$, the probabilities of finding the photon with correlated or anti-correlated MZ modes ($\ket{C}$ or $\ket{A}$) and correlated or anti-correlated I-I modes ($\ket{c}$  or $\ket{a}$) are: 
\begin{equation}
\begin{aligned}
    P_{C,c}&= \int dx\ |f_0(x)|^2 \cos^2\varphi_0 \cos^2(\theta x), \\
    P_{A,c}&= \int dx\ |f_0(x)|^2 \sin^2\varphi_0 \cos^2(\theta x), \\
    P_{C,a} &= \int dx \ |f_0(x)|^2 \sin^2\varphi_0 \sin^2(\theta x), \\
    P_{A,a} &= \int dx \ |f_0(x)|^2 \cos^2\varphi_0 \sin^2(\theta x).
\end{aligned}
\end{equation}

As before, we sum over the MZ interferometer modes $\ket{A}$ and $\ket{C}$ to find the probabilities 
\begin{equation}
\begin{aligned}
    P_{c}&= \int dx \ |f_0(x)|^2 \cos^2(\theta x), \\
    P_{a} &= \int dx \ |f_0(x)|^2 \sin^2(\theta x),
\end{aligned}
\end{equation}
which are independent of $\varphi_0$.  Likewise, $P_C=P_{C,c}+P_{C,a}$ and $P_A=P_{A,c}+P_{A,a}$ are independent of $\theta$, and consequently the Fisher information for measurement of $\theta$ and $\varphi_0$ are given by Eq. (\ref{eq-FI-2p}).  


\section{Conclusions}

It is not surprising that the quantum Cram\'er-Rao bound on the precision of  estimates of the phase gradient for a single-photon beam of finite width is consistent with the Heisenberg uncertainty principle. After all, the phase gradient corresponds to the transverse component of an optical field's wave vector. Nevertheless, it is interesting that this relation can be derived from the quantum Fisher information, thereby establishing a formal tradeoff between quantum sensitivity and spatial resolution. 
It is interesting, but also not unexpected, that the quantum precision limit for the two-photon state is superior by the ubiquitous factor of 2 for the maximally entangled state.  

It is remarkable that the quantum bound for estimation of the phase gradient may be attained by use of experimentally relevant binary projective measurements, similar to those used for phase measurement \cite{Tham2017,Tsang2016,Tang2016,Paur2016}, namely by use of an image-inversion interferometer instead of a Mach-Zehnder interferometer.  Although saturation of the quantum Fisher information for estimation of the phase gradient requires use of a narrow probe beam, this is in fact desirable since it enables greater resolution of scanning systems.  The sensitivity drops with increase of the beam width for both single-photon and two-photon probes, and the drop is faster for the two-photon state, so that its ubiquitous factor of 2 advantage is lost and even reversed for large beam width --- another manifestation of the fragility of quantum sensitivity.  



Another notable finding is that the precision bounds dictated by the quantum Fisher information for concurrent estimation of the phase \emph{and} the phase gradient are the same as for their independent estimation, for both single- and two-photon quantum states. The same conclusion applies to the structured interferometric configurations for which the multiparameter estimation is based on the classical Fisher information measure.  This was enabled by use of special combinations of a Mach-Zehnder interferometer and wavefront-division image-inversion interferometers producing measurement outcomes that are sensitive to only the phase and others that are sensitive to only the phase gradient.

\section*{Funding Information}
Air Force Research Laboratory grant  FA8651-19-2-0001

\bibliography{references}
\end{document}